\documentclass[sigconf]{acmart}
\usepackage[ruled,vlined,linesnumbered]{algorithm2e}
\usepackage{balance}

\usepackage{multirow}
\usepackage{booktabs}
\usepackage{enumitem} 
\usepackage{comment}
\usepackage{url}
\usepackage{caption}
\usepackage{subcaption}
\usepackage{xcolor}
\usepackage{footmisc}
\usepackage{array}
\colorlet{RED}{red}
\colorlet{OLIVE}{olive}
\usepackage{adjustbox}
\usepackage[utf8]{inputenc}
\setcopyright{none}
\renewcommand\footnotetextcopyrightpermission[1]{}
\newcolumntype{x}[1]{>{\centering\let\newline\\\arraybackslash\hspace{0pt}}p{#1}}

\definecolor{deepblue}{RGB}{0,66,66}

\makeatletter
\newcommand*{\rom}[1]{\expandafter\@slowromancap\romannumeral #1@}
\makeatother

\acmConference{Product Search}{Benchmark Paper}

\begin{document}

\title[Shopping Queries Dataset]{\emph{Shopping Queries} Dataset: A Large-Scale ESCI Benchmark for Improving Product Search}
\author{Chandan K. Reddy, Llu\'{\i}s M\`arquez, Fran Valero, Nikhil Rao, Hugo Zaragoza, Sambaran Bandyopadhyay, Arnab Biswas, Anlu Xing, Karthik Subbian}
\affiliation{Amazon \country{USA}}
\email{{ ckreddy,lluismv, fvalero, nikhilsr, hugzarag, sambarab, abisway, anluxing, ksubbian}@amazon.com}
\renewcommand{\shortauthors}{Reddy et al.}

\begin{abstract}
Improving the quality of search results can significantly enhance users experience and engagement with search engines. In spite of several recent advancements in the fields of machine learning and data mining, correctly classifying items for a particular user search query has been a long standing challenge, which still has a large room for improvement. This paper introduces the ``Shopping Queries Dataset'', a large dataset of difficult Amazon search queries and results, publicly released with the aim of fostering research in improving the quality of search results. The dataset contains around 130 thousand unique queries and 2.6 million manually labeled (query,product) relevance judgements. The dataset is multilingual with queries in English, Japanese, and Spanish. The Shopping Queries Dataset is being used in one of the KDDCup'22 challenges. 
In this paper, we describe the dataset and present three evaluation tasks along with baseline results: (i) ranking the results list, (ii) classifying product results into relevance categories, and (iii) identifying substitute products for a given query. We anticipate that this data will become the gold standard for future research in the topic of product search. 
\end{abstract}

\ccsdesc[500]{Information systems~Retrieval models and ranking}
\ccsdesc[300]{Information systems~Query representation}
\ccsdesc[500]{Applied computing~Online shopping}
\keywords{search relevance, querying, e-commerce, semantic matching}


\maketitle

\section{Introduction}

Improving the relevance of search results can significantly improve the customer experience and their engagement with search \cite{ANTHEM22}. Despite the recent advancements in the field of machine learning, correctly classifying items for a particular user search query for shopping is a challenging problem \cite{Choudhary22}. The presence of noisy information in the results, the difficulty of understanding the query intent, and the diversity of the items available are some of the reasons that contribute to the complexity of this problem.

When developing machine learning models for online shopping applications, extremely high accuracy in ranking is needed. This is even more stringent requirement when deploying search in mobile and voice search applications, where even a small number of irrelevant items can significantly hurt the user experience. Furthermore, the notion of binary relevance limits the customer experience. Specifically, classifying each product shown in response to a user query as being relevant or not typically yields results that have a detrimental effect on user experience. For example, for the query “iPhone”, would an iPhone charger be relevant, irrelevant, or somewhere in between? In fact, many users issue the query “iPhone” to find and purchase a charger for the iPhone. They simply expect the search engine to understand their need. For this reason, we break down relevance into the following four classes which are used to measure the relevance of items in the search results:
\begin{itemize} [leftmargin=*]

\item \textbf{Exact (E):} the item is relevant for the query, and satisfies all the query specifications (e.g., a water bottle matching all attributes of a query “plastic water bottle 24oz”, such as material and size)

\item \textbf{Substitute (S):} the item is somewhat relevant, i.e., it fails to fulfill some aspects of the query but the item can be used as a functional substitute (e.g., fleece for a “sweater” query)

\item \textbf{Complement (C):} the item does not fulfill the query, but could be used in combination with an exact item (e.g., track pants for “running shoes” query)

\item \textbf{Irrelevant (I):} the item is irrelevant, or it fails to fulfill a central aspect of the query (e.g., socks for a “telescope” query, or a wheat flour bread for a “gluten--free bread” query)
\end{itemize}

\begin{table*}
    \caption{Summary of the Shopping queries dataset for the task 1 (small version): the number of unique queries, the number of judgements, and the average number of judgements per query (Avg. Depth).}
\begin{adjustbox}{max totalsize={\textwidth}{\textheight},center}
\centering
    \begin{tabular}{l|ccc|ccc|ccc|}
    \cline{2-10}
     & \multicolumn{3}{c|}{\textbf{Total}} & \multicolumn{3}{c|}{\textbf{Train}} & \multicolumn{3}{c|}{\textbf{Public Test}} \\ \hline
    \multicolumn{1}{|l|}{\textbf{Language}} & \multicolumn{1}{l}{\textbf{\# Queries}} & \multicolumn{1}{l}{\textbf{\# Judgements}} & \multicolumn{1}{l|}{\textbf{Avg. Depth}} & \multicolumn{1}{l}{\textbf{\# Queries}} & \multicolumn{1}{l}{\textbf{\# Judgements}} & \multicolumn{1}{l|}{\textbf{Avg. Depth}} & \multicolumn{1}{l}{\textbf{\# Queries}} & \multicolumn{1}{l}{\textbf{\# Judgements}} & \multicolumn{1}{l|}{\textbf{Avg. Depth}} \\ \hline
    \multicolumn{1}{|l|}{English (US)} & 29,844 & 601,462 & 20.2 & 20,888 & 419,730 & 20.1 & 4,477 & 91,062 & 20.3\\
    \multicolumn{1}{|l|}{Spanish (ES)} & 8,049 & 218,826 & 27.2 & 5,632 & 152,917 & 27.2 & 1,208 & 32,905 & 27.2\\
    \multicolumn{1}{|l|}{Japanese (JP)} & 10,407 & 297,882 & 28.6 & 7,284 & 209,091 & 28.7 & 1,561 & 43,832 & 28.1\\ \hline
    \multicolumn{1}{|l|}{Overall} & 48,300 & 1,118,170 & 23.2 & 33,804 & 781,738 & 23.1 & 7,246 & 167,799 & 23.2 \\ \hline
    \end{tabular}
    \end{adjustbox}

    \label{tab:dataset-size-1}
    
\end{table*}

\begin{table*}
\caption{Summary of the Shopping queries dataset for the tasks 2 and 3 (large version): the number of unique queries, the number of judgements, and the average number of judgements per query (Avg. Depth).}
    
\begin{adjustbox}{max totalsize={\textwidth}{\textheight},center}
\centering
    \begin{tabular}{l|ccc|ccc|ccc|}
    \cline{2-10}
     & \multicolumn{3}{c|}{\textbf{Total}} & \multicolumn{3}{c|}{\textbf{Train}} & \multicolumn{3}{c|}{\textbf{Public Test}} \\ \hline
    \multicolumn{1}{|l|}{\textbf{Language}} & \multicolumn{1}{l}{\textbf{\# Queries}} & \multicolumn{1}{l}{\textbf{\# Judgements}} & \multicolumn{1}{l|}{\textbf{Avg. Depth}} & \multicolumn{1}{l}{\textbf{\# Queries}} & \multicolumn{1}{l}{\textbf{\# Judgements}} & \multicolumn{1}{l|}{\textbf{Avg. Depth}} & \multicolumn{1}{l}{\textbf{\# Queries}} & \multicolumn{1}{l}{\textbf{\# Judgements}} & \multicolumn{1}{l|}{\textbf{Avg. Depth}} \\ \hline
    \multicolumn{1}{|l|}{English (US)} & 97,345 & 1,819,105 & 18.7 & 68,139 & 1,272,626 & 18.7 & 14,602 & 274,261 & 18.8\\
    \multicolumn{1}{|l|}{Spanish (ES)} & 15,180 & 356,578 & 23.5 & 10,624 & 249,721 & 23.5 & 2,277 & 53,494 & 23.5 \\
    \multicolumn{1}{|l|}{Japanese (JP)} & 18,127 & 446,055 & 24.6 & 12,687 & 312,397 & 24.6 & 2,719 & 66,612 & 24.5\\ \hline
    \multicolumn{1}{|l|}{Overall} & 130,652 & 2,621,738 & 20.1 & 91,450 & 1,834,744 & 20.1 & 19,598 & 394,367 & 20.1\\ \hline
    \end{tabular}
    \end{adjustbox}
    \label{tab:dataset-size-2}
    
\end{table*}

\begin{table*}
    \caption{ESCI distribution (in \%) of the Shopping queries dataset.}
    \begin{adjustbox}{max totalsize={\textwidth}{\textheight},center}
    \centering
        \begin{tabular}{l|crrr|crrr|crrr|}
            \cline{2-13}
            & \multicolumn{4}{c|}{\textbf{Total}} & \multicolumn{4}{c|}{\textbf{Train}} & \multicolumn{4}{c|}{\textbf{Public Test}} \\ \hline
            \multicolumn{1}{|l|}{\textbf{Dataset version}} & \textbf{E} & \multicolumn{1}{c}{\textbf{S}} & \multicolumn{1}{c}{\textbf{C}} & \multicolumn{1}{c|}{\textbf{I}} & \textbf{E}                  & \multicolumn{1}{c}{\textbf{S}} & \multicolumn{1}{c}{\textbf{C}} & \multicolumn{1}{c|}{\textbf{I}} & \textbf{E}                  & \multicolumn{1}{c}{\textbf{S}} & \multicolumn{1}{c}{\textbf{C}} & \multicolumn{1}{c|}{\textbf{I}} \\ \hline
            \multicolumn{1}{|l|}{\textbf{Small}} & \multicolumn{1}{r}{43.72} & 34.33 & 5.13 & 16.82 & \multicolumn{1}{r}{43.64} & 34.28 & 5.19 & 16.89 & \multicolumn{1}{r}{44.06} & 34.59 & 4.87 & 16.48 \\
            \multicolumn{1}{|l|}{\textbf{Large}} & \multicolumn{1}{r}{65.20} & 21.91 & 2.89 & 10.00 & \multicolumn{1}{r}{65.21} & 21.89 & 2.91 & 10.00 & \multicolumn{1}{r}{65.16} & 22.02 & 2.84 & 9.99 \\ \hline
        \end{tabular}
    \end{adjustbox}
\label{tab:dataset-esci}
    
\end{table*}

In this paper, we introduce the “Shopping Queries Dataset”, a large dataset of difficult search queries published with the aim of fostering research in the area of semantic matching of queries and products. For each query, the dataset provides a list of up to 40 results, together with their ESCI relevance judgements (Exact, Substitute, Complement, or Irrelevant) indicating the relevance of the product to the query \cite{mcauley2015inferring}. Each query-product pair is accompanied by additional information from the Amazon catalog, including: product title, product description, and additional product related bullet points. This information is public, as it is displayed at the Amazon website when searching for those products. The Shopping Queries Dataset is multilingual, as it contains queries in English, Japanese, and Spanish \cite{ahuja2020language}. With this data, we propose three different tasks, consisting of: 1) ranking the results list, 2) classifying the query/product pairs into E, S, C, or I categories, and 3) identifying substitute products for a given query.

This collection has some characteristics which we think make it specially interesting for ML research in retrieval and classification. First, it is derived from real customers searching for real products online. Second, it provides both breadth (a large number of queries, in three languages) and depth ($\approx$20 results per query), unlike other existing large document retrieval collections which tend to provide either breadth or depth but not both (e.g. MSMarco \cite{MsMarco}, TRECDL \cite{TRECDeep21}, NLQEC \cite{Papenmeier2021DatasetON}). Third, all results have been manually labeled with multi-valued relevance labels, describing difference relevance status (in the context of e-shopping). Fourth, queries have not been randomly sampled, but rather, subsets of the queries have been sampled specifically to provide a variety of challenging problems (such as negation, attribute parsing, etc.). Fifth, we also provide descriptions of the retrieval objects which have categorical and textual metadata, as well as multiple levels of representation (from a short title to a long description of the product). Finally, products are linked to the online Amazon catalog.

\section{Shopping Queries Dataset}
\label{sec:dataset}
The Shopping Queries Dataset is a multilingual large-scale manually annotated dataset composed of challenging customer queries. 
The training dataset contains a list of query-result pairs with annotated E/S/C/I labels. The data is multilingual, and it includes queries from English, Japanese, and Spanish languages.

This dataset can serve as a standard benchmark for building algorithms to measure and improve search quality in the years to come. The code for the baseline methods is made available at \footnote{\href{https://github.com/amazon-research/esci-code}{https://github.com/amazon-research/esci-code}}. Note that the complete dataset will be made available at this repository after the KDDCup competition is completed.

\subsection{Query Selection Process}
Most frequent shopping queries are easy to solve with standard state-of-the-art search engine techniques and result in near-perfect results. For this reason, a randomly sampled query set would not be of great interest to the research community. Hence, we had to develop a methodology to sample \textit{interesting} or \textit{challenging} queries, a difficult and open problem in itself. To achieve this goal, we explored various sampling strategies to select the ones returning more mistakes in several of our production baseline models. While this approach is clearly biased by the selection of baseline models, in our experience, it leads to results that are  more interesting (i.e., harder for many models) than those obtained by using sampling methods directly tied to the properties of a particular model (such as those obtained by active learning or adversarial approaches used to improve model training such as \cite{bilgic:sigir12, WinoGrande}). In the rest of this section, we provide a brief summary of the selected strategies.

\begin{description}
   \item[Behavioral] We use several statistics to sample queries leading to results or purchases with non-representative click distributions.
\item[Negations] We use several regular expressions to sample queries with negations. (for e.g., `energy bar without nuts'.)
\item[Parse Pattern] We use several regular expressions on the parsed query to sample queries with some linguistic complexity, such as queries containing quantities, a product type with an adjective, etc.  (for e.g., 'gluten free english biscuits'.)
\item[Price Pattern] We use several statistics to sample queries leading to results or purchases with non-representative price distributions.
\item[Other] We sample queries from a number of random query sampling processes, removing those that result in perfect or near perfect results.
\item[NLQEC] Queries from the NLQEC dataset \cite{Papenmeier2021DatasetON} with 30 tokens or less.
\end{description}

\subsection{Annotation of Relevance Judgements}
Each query--result pair was manually annotated with E/S/C/I labels by humans trained on the task. A minimum of three annotations were collected for each pair, and an automatic aggregation mechanism selected the majority vote as the gold label. Each language had a different pool of annotators. The information that the annotators could see to assign the output labels was the detail page of the product, as it appears on www.amazon.com website. Since there is human annotation involved in the process, the resulting labels are not perfect and can be noisy. We estimated the accuracy of the human annotations by randomly sampling 100 cases per language and carefully inspecting the assigned labels. The overall agreement is 91\%. The majority of the discrepancies ($\approx$50\% of them) are Irrelevant cases that the judges considered valid Substitutes. We observed some bias in the annotators to be less strict when applying the definition of Substitute. Only a small percentage of discrepancies ($\approx$15\%) correspond to extreme cases (for e.g., confusions between Exact and Irrelevant), which can be attributed to annotation mistakes. We know that fine distinction between E/S/C/I labels can be difficult in cases when the query is ambiguous or not well specified, or when it comes to determine whether some aspect of the product not matching the query is fundamental or not for the customer (i.e., distinguishing between Substitute and Irrelevant). If we consider a simpler binary distinction between Exact and Not-Exact (grouping together Substitute, Complement, and Irrelevant) labels, the agreement between our judgement and that from the manual annotators increases to $>$96\%.  

\subsection{Product Description}
Every example in the dataset contains a query--result pair, the gold E/S/C/I label, and a number of fields that can be used as features to train classification or ranking models. More concretely, the list of fields (columns) for each entry are (in this order):
\begin{itemize}
    \item example\_id: example identifier 
    \item query: text string representing the customer query
    \item query\_id: a unique identifier of the query
    \item product\_id: product identifier, which references a specific product in the amazon.com site
    \item product\_title: the title of the product as it appears in the amazon.com site  
    \item product\_description: the product description field as it appears in the amazon.com site  
    \item product\_bullet\_point: the bullet point descriptions of the product as it appears in the amazon.com site  
    \item product\_brand: string representing the brand of the product
    \item product\_color: string representing the color of the product, if applicable
    \item product\_locale: the locale from which the product is selected (either US, Spain, or Japan)
    \item esci\_label: the output label to be predicted (E, S, C, or I)
\end{itemize}

\subsection{Splits and Statistics}
The data is stratified by queries in three splits: \textit{training}, \textit{public} test, and \textit{private} test, at 70\%, 15\%, and 15\%, respectively.  
We propose three different tasks on the dataset, as described in Section~\ref{tasks}, which are precisely the ones from the Amazon KDD Cup '22.\footnote{https://www.aicrowd.com/challenges/esci-challenge-for-improving-product-search} The training split is intended for training the classification and ranking systems, the public test is to be used to tune the models, and the private test is a held-out split to be used to evaluate only the results.

We provide two different versions of the dataset, one for Task1 (query-product ranking) and another which is a superset of the first for Tasks 2 and 3 (multiclass product classification, and product substitute identification). The larger version of the dataset contains 130,652 unique queries and 2,621,738 judgements, corresponding each to a (query--result) judgement. The smaller version of the dataset contains 48,300 unique queries and 1,118,117 rows. The smaller version is a subset of the larger version where the simpler queries (in terms of NDCG) are filtered out.

A summary of the Shopping Queries Dataset is given in the Tables~\ref{tab:dataset-size-1} and~\ref{tab:dataset-size-2} showing the statistics of the small and large version, respectively. These tables include the number of unique queries, the number of judgements, and the average number of judgements per query (i.e., average depth) across the three different locales (languages). Table ~\ref{tab:dataset-esci} shows the ESCI distribution of these two versions of the dataset for only  the training and public test set.\footnote{We omit any information on the private test set, in the current version of this paper, as the competition is ongoing at the time of writing it.} The class labels are imbalanced, with label E being the most frequent, and label C being the least.

Across languages, we can see that the proportion of English queries (61.8\% and 74.5\% for the small and large versions, respectively ) is significantly larger than the proportion of Spanish (16.6\% and 11.6\%) and Japanese (21.5\% and 13.9\%) queries. On the contrary, the average number of results per query in English is slightly smaller than in Spanish and Japanese, which are comparable. 

Finally, we can observe that the larger version is an ``easier'' dataset than the smaller version, as the proportion of Exact matches is higher (65.2\% vs 43.7\%) and the number of results per query is smaller (20.1 vs 23.2).  

\section{Tasks and Evaluation Metrics}\label{tasks}
This dataset can aid in building new ranking strategies and simultaneously identify interesting categories of results (i.e., substitutes) that can be used to improve the customer experience when searching for products. Some of the potential tasks that can performed using our Shopping Queries Dataset are:

\begin{enumerate} [leftmargin=*]
\item Query-Product Ranking
\item Multiclass Product Classification
\item Product Substitute Identification
\end{enumerate} 

\subsection{Task 1: Query-Product Ranking}
Given a user specified query and a list of matched products, the goal of this task is to rank the products so that the relevant products are ranked above the non-relevant ones. This is similar to standard information retrieval tasks, but specifically in the context of product search in e-commerce. The input to this task is a list of queries and for each query a list of products, with no specific order. The maximum number of products per query is 40, and at least one is guaranteed to be non irrelevant (either Exact, Substitute, or Complement). The products are described by the features explained in Section~\ref{sec:dataset}. The goal is to sort for every query the list of products in decreasing order of relevance, i.e., first the Exact matches, then Substitutes, followed by Complements, and Irrelevants at the end.

The task performance will be evaluated using Normalized Discounted Cumulative Gain (nDCG)\footnote{\href{https://en.wikipedia.org/wiki/Discounted_cumulative_gain}{https://en.wikipedia.org/wiki/Discounted\_cumulative\_gain}}~\cite{jarvelin-kekalainen-02}. 
This is a commonly used relevance metric in the literature. 
Highly-relevant documents appearing lower in a search results list should be penalized as the graded relevance is reduced logarithmically proportional to the position of the result. In our case, we have 4 degrees of relevance for each query and product pair: Exact, Substitute, Complement, and Irrelevant, and we set gain values of 1.0, 0.1, 0.01, and 0.0, respectively.  
Note that there is a corner case where nDCG is not well defined, i.e., when all results are Irrelevant. This is not possible in our case since all queries have at least one non irrelevant result.




\subsection{Task 2: Multiclass Product Classification}

Given a query and a result list of products retrieved for this query, the goal of this task is to classify each product as being an Exact, Substitute, Complement, or Irrelevant match for the query. This is a multi-class classification problem.
The input is a list of <query,product> pairs, along with all the product features described in Section~\ref{sec:dataset}. The output is a class label for each of the input pairs. 
We use F1 score\footnote{\href{https://en.wikipedia.org/wiki/F-score}{https://en.wikipedia.org/wiki/F-score}}~\cite{van-rijsbergen-79} to evaluate performance of this task. We decided to use the micro-averaged version across classes, because the four classes are unbalanced (65.20\% Exacts, 21.91\% Substitutes, 2.89\% Complements and 10.00\% Irrelevants) and this metric is robust for this situation. 


\subsection{Task 3: Product Substitute Identification}
This task will measure the ability of the systems to identify the substitute products in the list of results for a given query. The notion of “substitute” is exactly the same as in Task 2. This is a binary classification task. For each $\langle$ query,product $\rangle$ input pair, the goal is to assign an output label "Substitute" or "non-Substitute". Since the goal is to identify positive substitute cases, we will use the F1 score to evaluate the results.  


\section{Experimental Results}\label{results}
In this section, we present a first exploration of the tasks defined on the Shopping Queries Dataset by evaluating standard ranking and classification approaches as baselines.

\subsection{Baseline Models}
For the first task (query-product ranking), we propose 
to use the MS MARCO Cross-Encoder Information retrieval model\footnote{\href{https://huggingface.co/cross-encoder/ms-marco-MiniLM-L-12-v2}{https://huggingface.co/cross-encoder/ms-marco-MiniLM-L-12-v2}}~\cite{reimers2019sentence,nguyen2016ms} that encodes the query and product titles. We fine-tune the model on the US part of the training set using the default hyper-parameter configuration\footnote{\href{https://github.com/UKPLab/sentence-transformers/blob/master/examples/training/ms\_marco/train\_cross-encoder\_kd.py}{https://github.com/UKPLab/sentence-transformers/blob/master/examples/ training/ms\_marco/train\_cross-encoder\_kd.py}}, where we set the following parameters for the Cross-Encoder model:
maximum length=512, activation function=identity, and number of labels=1 (binary task). For the training hyperparameter configuration, we use MSE loss function, evaluation steps=5000, warm-up steps=5000, learning rate=7e-6, training epochs=1, and number of development queries=400. 

For the JP and ES locales, we propose to fine-tune two semantic search models, one for each locale based on a multilingual MPNet   model\footnote{\href{https://huggingface.co/sentence-transformers/all-mpnet-base-v1}{https://huggingface.co/sentence-transformers/all-mpnet-base-v1}}~\cite{song2020mpnet} that will also map the queries and product title. We also use the default hyper-parameter configuration\footnote{\href{https://www.sbert.net/docs/training/overview.html}{https://www.sbert.net/docs/training/overview.html}}, with 
cosine similarity as the loss function, one training epoch, 100 evaluation steps, and 200 development queries. For all the locales during training, we map the exact labels to 1.0 and the other labels (substitute, complement and irrelevant) to 0.0 as a binary task. In addition to this neural approach, we also experimented with the open search engine Terrie v5.5\footnote{\href{https://github.com/terrier-org/terrier-core}{https://github.com/terrier-org/terrier-core}}~\cite{macdonald2012puppy}, to index the entire product catalog considering the product title. To rank the results, we used the conventional BM25 model for all three locales together. 

\begin{figure}
    \includegraphics[width=0.3\textwidth]{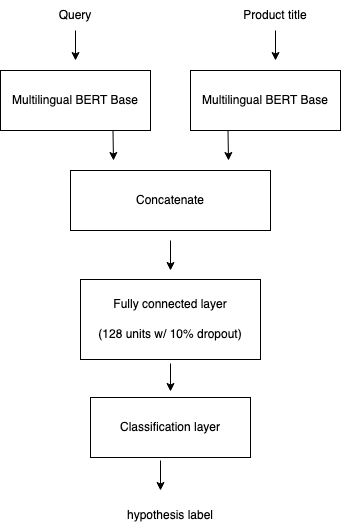}
    \caption{MLP classifier whose input is the concatenation of the representations provided by BERT multilingual base for the query and title of the product.}
   \label{fig:mlp-cls}
\end{figure}

For the other two tasks, multiclass product classification and product substitute identification, we develop a Multilayer Perceptron (MLP) classifier, whose input is the concatenation of the representations provided by BERT multilingual base model\footnote{\href{https://huggingface.co/bert-base-multilingual-uncased}{https://huggingface.co/bert-base-multilingual-uncased}}~\cite{devlin2018bert} for the query and title of the product. In this approach, the BERT representations are frozen. This approach performs the following steps (see Figure~\ref{fig:mlp-cls}): (1) calculate BERT representation of the query; (2) calculate BERT representation for product title; (3) concatenate BERT representation of query and product title; (4) apply a fully connected layer of 128 neurons (with 10\% dropout during training) [\textit{trained weights}]; (5) apply a classification layer [\textit{trained weights}]. We applied max--pooling to the BERT representations instead of using the representation of the \texttt{[CLS]} token. For the training hyper-parameter configuration, we set 4 epochs and Adam optimizer~\cite{kingma2014adam}, with values for epsilon, learning rate and weight decay of 1e-8 , 5e-5 and 0.01, respectively.  

\subsection{Results}

\begin{table*}[t]
    \caption{nDCG and Micro F1 baselines scores on the public test for all the tasks.}
    \label{tab:baselines-results}
    \begin{adjustbox}{max totalsize={\textwidth}{\textheight},center}
    \centering
    \begin{tabular}{|c|l|l|r|r|r|r|r|}
        \hline
        \textbf{Task} & \textbf{Model} & \textbf{Metric} & \textbf{Public Test} & \textbf{English}  & \textbf{Spanish} & \textbf{Japanese}\\\hline
        \multirow{2}{*}{\textbf{T1}}  & BM25 (all locales together) & \multirow{2}{*}{nDCG} & 0.563 & 0.675 & 0.697 & 0.136 \\
        & Fine-Tuned Cross-Encoder (EN) and MPNet (ES,JP) &  & 0.852 & 0.857 & 0.849 & 0.840\\ \hline
        \textbf{T2} & Frozen BERT MLP Classifier & Micro F1 & 0.656 & 0.685 & 0.580 & 0.595\\ \hline
        \textbf{T3} & Frozen BERT MLP Classifier & Micro F1 & 0.780 & 0.795 & 0.757 & 0.737\\ \hline
    \end{tabular}
    \end{adjustbox}
\end{table*}

Table ~\ref{tab:baselines-results} shows the results of the baselines in the public test set for the three tasks. 
Results are also presented broken down by language (English, Spanish, Japanese) corresponding to the US, ES and JP locales.

For Task 1, we can see that the neural approach gets a much better nDCG results than the Terrie-BM25 counterpart (0.852 vs. 0.551). It should be noted that this is not a fair comparison since we used the default configuration for the Terrie-BM25 approach, which obtains poor results on Japanese (due to non-Japanese specific pre-processing), thus significantly penalizing the averaged metric across the overall data. The neural approach obtains nDCG scores in the interval (0.840, 0.857), which are comparable across languages.

For the other two classification tasks, the BERT-based MLP classifier obtains results that are clearly better for English than for Spanish and Japanese (e.g., for Task 2, compare the F1 for English and Spanish, 0.685 vs. 0.580). In the future, we will investigate the reason(s) for these substantial differences. In Task 3, the results are still favorable to English but with smaller differences, half the size compared to the difference seen in Task 2. 

In all three tasks, but especially on classification tasks 2 and 3, we can see that there is a large room for improvement on top of the baseline models.

\section{Conclusion}\label{results}
In this paper, we introduce the Shopping Queries Dataset, a large--scale benchmark to improve the state-of-the-art algorithms for e-commerce product search. We first provide details about the dataset content and its main statistics. Then, we explain three evaluation tasks, which are the ones proposed in the KDDCup'22 challenge, and we provide some initial results from various baselines for reference. We hope the release of this dataset will spur research from the machine learning and data mining communities into developing scalable and high performing models for product search. 

\balance
\bibliographystyle{ACM-Reference-Format}
\bibliography{esci}

\end{document}